\documentclass[aps]{revtex4}
\usepackage{amsmath,amssymb}
\usepackage{color,graphicx}
\usepackage{multirow}
\usepackage{amsmath,amssymb}
\newcommand{\be}{\begin{equation}}
\newcommand{\ee}{\end{equation}}
\begin{document}
 \title{Killing (absorption) versus survival in random motion}
 \author{Piotr Garbaczewski}
 \affiliation{Institute of Physics, University of Opole, 45-052 Opole, Poland}
 \date{\today }
 \begin{abstract}
  We  address    diffusion processes in a bounded domain, while focusing on somewhat unexplored affinities  between
   the presence  of    absorbing  and/or   inaccessible  boundaries. For the  Brownian motion  (L\'{e}vy-stable cases
   are briefly mentioned)  model-independent  features   are established, of the dynamical law  that  underlies
     the short  time    behavior of  these  random  paths, whose overall life-time is predefined to be long.
       As a by-product, the  limiting  regime   of a permanent trapping in a domain  is obtained.
      We  demonstrate that the  adopted  conditioning method, involving the so-called   Bernstein transition   function,
        works properly  also  in an  unbounded domain,  for stochastic processes with killing
       (Feynman-Kac kernels play the role of transition densities),  provided   the spectrum  of the related semigroup operator
        is discrete.   The method is shown to   be useful in the case,  when the spectrum   of the generator goes down to
        zero and no   isolated minimal (ground state) eigenvalue is in existence, like e.g. in the problem of the   long-term
         survival on a half-line with  a sink at origin.
  \end{abstract}
 \maketitle

  \section{Motivation.}

Diffusion processes   in a bounded domain  (likewise the jump-type L\'evy processes)   serve as
important model systems  in the description of   varied spatio-temporal phenomena of  random origin in Nature.
When arbitrary  domain shapes are considered, one deals with  highly sophisticated problems on their own,
 an object of  extensive investigations in the mathematical literature.

A standard physical inventory, in  case of absorbing boundary conditions  which are our concern in the present paper, refers
  mostly to  the statistics of exits, e.g. first and mean first  exit times, probability of survival
and its asymptotic decay,  thence various aspects of the lifetime of the pertinent stochastic process in a  bounded domain,
 \cite{redner}-\cite{pekalski}, see also \cite{li,castro,lukic}.

Typically one interprets the survival probability as   the probability that not a single particle may hit
 the domain boundary before  a given time $T$.
The long-time survival is definitely not a property of the free Brownian motion in a domain with absorbing boundaries, where
the survival probability is known  to decay to zero  exponentially  with $T\rightarrow \infty $, \cite{redner0,agranov}.
 Therefore, the physical conditions   that ultimately give rise to a long-living random system, like e.g. those considered in
 \cite{agranov,redner0}, see also \cite{pekalski},   must result in a  specific remodeling  (conditioning, deformation,  emergent or
  "engineered"  drift)  of the "plain"  Brownian motion.

For simple geometries (interval, disk and the sphere) the exponential decay of the single-particle
survival probability has been identified   to  scale the stationary  (most of the time) gas density profile,
while that profile and  the  decay rates   stem directly from spectral solutions of the  related eigenvalue
 problem for the Laplacian with the Dirichlet boundary data, respectively  on the interval, disk and sphere, see e.g. \cite{agranov,redner0,redner}.
  In fact, the square of the  lowest eigenfunction, upon  normalization,  has been found to play the role of the pertinent  gas profile
  density, while the   associated  lowest eigenvalue  of the motion generator  determines the decay rate.

   These observations  have been established within so called  macroscopic fluctuation theory (of particle survival).
      Effectively, that was  also  the case in Ref. \cite{pekalski}, where a suitable choice of the Monte Carlo
  updating procedure   has resulted in the  increase of the survival probability in the diffusion
model.  That has been paralleled by  a temporarily favored motion  trend (engineered drift) -
 away from the boundaries, directed towards the mid-interval  locations.

The long lifetime  regime   of  a diffusion process  in  a bounded domain  may be comparatively set against that  the infinite lifetime
       (trap, with  the  inaccessible boundaries) see e.g.    \cite{gar,gar1}  and references
     therein.  The limiting  behavior  (with respect to the  life-time)  of absorbing diffusion processes and  symptoms of  their convergence
      towards   permanently trapped   relatives (never leaving a bounded domain)   is worth investigation.

It is useful to mention  our earlier  analysis of  the permanent
trapping problem (including a fairly serious  question about what is actually meant by  the Brownian motion
in a trap (interval), with a preliminary discussion of that issue for  jump-type processes of L\'{e}vy type, \cite{gar1}.

A common formal  input, both for absorbing and permanently trapped  diffusion processes in a bounded domain,  is that of  spectral   problems
 for Dirichlet Laplacians  and   Laplacians  perturbed by suitable potentials.  c.f. \cite{gar,gar1,vilela,vilela1}.
The notion of Markovian semigroup operators and their integral (heat) kernels is here implicit and  a path-wise description by means
 of the Feynman-Kac formula  is feasible.

That entails an exploration of   affinities  of  general  killed   diffusion
processes with  diffusions    with  an infinite life-time. We point out that the notion of killing stems from a probabilistic
 interpretation of  the  Feynman-Kac formula, \cite{roepstorff,faris,davies}.

Our  discussion departs from  much   earlier investigations  of random processes that either stay  forever within a prescribed convex domain
or  are bound to  avoid such a domain while "living"  in its exterior, c.f. \cite{redner}. We are strongly      motivated by the
past mathematical research,  whose roots can  be traced back to   Knight's "taboo processes", \cite{knight,pinsky,korzeniowski}.

We  introduce  a direct  conditioning  method, that   essentially  relies on path-wise intuitions  underlying   the notion of the
  Feynman-Kac semigroup transition kernel, given the  diffusion generator.
 It is  based on the concept of the  Bernstein   transition  function (actually a conditional probability  density), \cite{jamison,zambrini}, which
 in the present paper is  explored as a diagnostic tool for  the description of   dynamical  properties, e.g. an  emergent  dynamical law,
 of short-time  paths  segments in  a "bunch", of pre-defined to be   long-living, sample trajectories of a  stochastic diffusion process with absorption.

   The Bernstein function has an appearance of  the $(x,t)$-dependent probability density,  associated with  paths pinned at  two
    fixed  space-time poinnts: $(y,s)$ (initial) and $u,T$ (terminal), $s<t<T$.  Our finding is that  if  $T\gg t$ is large enough,  then
    the Bernstein function is approximated by (and, with $T\rightarrow \infty $, ultimately reduces to)
    the transition probability density  $p(x,t|y,s)$ of the diffusion process  with an infinite life-time, where the $u$ dependence is absent.
    The is the embodiment of the sought for dynamical law.   The corresponding Fokker-Planck equation follows.

  Our  major  playground  are Markovian  diffusion processes in a bounded domain with  absorbing  boundaries.
 Next, we   shall   demonstrate tha validity of  extensions of our  strategy   to  absorbing processes  in unbounded domains (like e.g. sink on the  half-line)
  and  to  mmore general  Markovian processes with killing.

  We indicate that our    conditioning  method  is not specific to the  standard Brownian (diffusions)   "territory".  One
  can  readily  pass  from Brownian diffusions to jump-type   L\'{e}vy stable
    stochastic processes, whose restriction to  a bounded domain, or the case of an unbounded domain with a sink, have been an object of
    vivid studies in the current literature.

  \section{Interval with absorbing ends and terms  of  survival.}

\subsection{Preliminaries.}

Diffusion processes in the interval with various boundary  conditions have become favored model systems in the
 statistical physics  approach to the  Brownian motion  and  were often  exploited in unexpected settings (prisoner in an expanding cage on the  cliff),
 including the   extensions of the formalism to higher dimensions, \cite{risken,redner},
 see also \cite{redner1,redner2,li}.  In Ref. \cite{redner} there is the whole chapter about diffusion
  in the interval with absorption at its ends.  Other  inspirations  (bounded variation of interest in economy, harmonic and optical traps)
    can be gained from \cite{li,castro,lukic}, see also \cite{gar1} where the permanent trapping problem has been addressed. \\

Let us consider the free   diffusion      $\partial _t k= D \Delta _xk $  within an interval ${\cal{D}} = [a,b] \subset  R$,
 with absorbing boundary conditions at its end points  $a$ and $b$.
 The   time and space homogenous  transition density,   with $x, y  \in  (a,b)$, $0\leq s<t$ and $b-a=L$ reads

\be
k(x,t|y,s) ={\frac{2}L} \sum_{n=1}^{\infty } \sin\left( {\frac{n\pi
}L}(x - a)\right) \sin\left( {\frac{n \pi }L} (y - a)\right)  \exp
\left(- {\frac{Dn^2\pi ^2}{L^2}} (t-s) \right).
\ee

Note  that  $\lim_{t\rightarrow s} k(x,t|y,s) \equiv \delta(x-y)$.
 We point out that in view of the time homogeneity, we can write $k(x,t|y,s)=k_{t-s}(x|y)$.   One should keep in memory
 that $k$ is a  symmetric function of  $x$ and $y$, i.e.  $k(yt|x,s)=k(y,t|x,s)$.

We deliberately use the notation $k(x,t|y,s)$ if probability is not conserved by the dynamics (it "leaks out" and decays with the growth of time)
  which is the case  for absorbing boundary conditions.
   If the transition density would  conserve probability in the interval (that  corresponds to a diffusion never leaving ${\cal{D}} $),
    the standard notation  $p(x,t|y,s)$ will be used.

While setting $a=0$, $s=0$, and $y=x_0$  in (1) we arrive at a
customary notion  of the concentration $k(x,t|x_0,0)= c(x,t)$, typically   employed in the literature, c.f. \cite{redner}.
Initially  one has $\int _0^L c(x,t=0)\, dx =1$, hence  for $t>0$
the absorption at the boundaries  enforces a  decay of the probability density, whose time  rate may be quantified by
 $(d/dt) \int _0^L c(x,t) dx $ where $t>0$.  The decay  of $c(x,t)$ is known to be exponential and its time rate is determined by
  the lowest positive eigenvalue of the Laplacian in a bounded domain (the role of so called eigenfunction expansions  \cite{risken}
  needs to be emphasized).

  To simplify notation, we note  that   $x\rightarrow  x'= (x-a)/L$ transforms  the
interval $[a,b]$ into $[0,1]$.  Another transformation $x\rightarrow x'=x - {\frac{1}2} (a+b)$ maps  $[a,b]$ into $[-c,c]$, with $c=L/2$,
 whose special case (set $L=2$) is the interval $[-1,1]$.     From now on we employ the symmetric interval $[-c, c]$ with
$L=2c, c>0$.   We also  set  $D=1/2$.

\subsection{(Dirichlet)  heat kernel in a bounded domain and its path-wise interpretation.}

  The   transition density $ k(x,t|y,s)$,  as   defined by  Eq. (1),    is known to be an  integral (heat) kernel of the semigroup operator
 $\exp [{\frac{1}2}(t-s) \Delta _{{\cal{D}}}]$,
  where  the notation $\Delta _{{\cal{D}}}$ directly  refers to the  standard Laplacian in the domain  ${\cal{D}}$, with absorbing boundary conditions.
The pertinent diffusion process   is Markovian and has an  explicit  semigroup
encoding in terms of the motion operator  $T_{t-s} = \exp [{\frac{1}2} \Delta _{\cal{D}}(t-s)]$.

The semigroup property $T_{t-s}  T_{s-r} =T_{t-r}$, with  $t>s>r$, implies
the validity of  the composition rule
$$\int_{\cal{D}} k(x,t|y,s)\, k(y,s|z,r)\, dy = k(x,t|z,r).$$
  Given a suitable function   $f(x)$. Its semigroup evolution  is defined as follows
$$f(x) \rightarrow  f(x,t)= (T_t f)(x) =  \exp [{\frac{1}2} \Delta _{\cal{D}} t] f(x) = \int_{\cal{D} } k(x,t|y,0) f(y) dy.
 $$
and $f(x,t)$ is a   solution of the  "heat" equation  in ${\cal{D}}$:   $\partial _t f(x,t) ={\frac{1}2} \Delta _{\cal{D}}  f(x,t)$.

Semigroup kernels admit the  paths-wise interpretation by means  of the Feynman-Kac formula  (e.g. the path integral),  which
has become a classic  \cite{roepstorff} (actually  that is possible in any   bounded domain in $R^n$,
  an interval being a particular  case).
  Indeed,  $k(x,t|y,s)$    is  prescribed for a  specific  "bunch"  $\omega \in \Omega _{y,s}^{x,t}$    of   sample paths of the Brownian motion:
   $\omega (\tau), s\leq \tau \leq t$  (here $\omega (s)=y$  stands for the point of origin
  while  $\omega (t)=x$)   is the target/destination  point), all of    which  survive within  $ {\cal{D}} \equiv (-c,c)$ up to time $t>s$, e.g. their
  killing time   exceeds $t$. We have:
  \be
 [\exp({\frac{1}2}  \Delta _{{\cal{D}}}\, (t-s))](x,y)  = k(x,t|y,s)=
\int d\mu _{y,s}^{x,t}(\omega) = \mu _{y,s}^{x,t}(\omega).
 \ee
The notion of the conditional Wiener measure of the set  of random  paths connecting $y$ with $x$, in the time interval $(s,t)$ is here  implicit,
 \cite{roepstorff}.  (Note that the term  "conditional" means here  that  two endpoints $y$ and $x$  are fixed, while in case of the Wiener measure
  only  one  point, that of   the origin of random motion, is fixed.).

    Following the traditional lore, we say that  $\mu _{y,s}^{x,t}(\omega)$ is a total mass
  of the pinned paths set  $\Omega _{y,s}^{x,t}$.  It is the transition kernel $k(x,t|y,s)$ which is the  pertinent mass measure.

For clarity of arguments, we  presume   $y<x$). Let us choose  a  window $I= [a,b] \subset  [y,x]$   with $a<b$,  (it is a standard  preparatory step in the construction of the conditional Wiener measure).
  We can  assign a  mass to a specific subset  $\Omega _{y,s}^{x,t}(I)$   of the considered  pinned  paths  set, that comprises these
   sample paths only, which at  an intermediate time $r, s<r<t$  reach/cross  the window $I$, before approaching
 the  final  destination $x$ at time $t$.  The  mass  of  such subset of sample paths is known to  be:
 \be
\mu _{y,s}^{x,t}(I) = \int _I  k(x,t|u,r)k(u,r|y,s) du \,  < \,  \mu _{y,s}^{x,t}({\cal{D}}) = k(x,t|y,s)
\ee
Note that with $I$ replaced by ${\cal{D}}$ in the above integral, the semigroup composition rule  yields the mass   $k(x,t|y,s)$.

The transition density $k(x,t|y,s)$ allows us to define a  probability that the process started at $y$ at time $s$, will actually reach an interval $(x,x+\Delta x)$
at time $t>s$. It reads  $P(\Omega ([y,x]),\Delta x)= k(x,t|y,s) \Delta x$. Likewise we obtain a probability
$P(\Omega (I), \Delta x) =\mu _{y,s}^{x,t}(I)  \Delta x$   for the
 $I$-constrained subset of paths.

The ratio
 \be
 {\frac{P(\Omega (I), \Delta x)} {P(\Omega ([y,x]),\Delta x)}} = \int_I  \frac{ k(x,t|u,r)\, k(u,r|y,s)}{{k(x,t|y,s)}}  du
\ee
is nothing but a conditional probability  quantifying the fraction of mass of the  a subset of paths crossing  $I$ at time $r$,  while set against
 the overall mass of all sample paths  with origin  $y$ at $s$ and destination   $x$,  at  time  $t$, $s<r<t$.

     Under the integral sign  in Eq.  (4) we encounter a conditional  probability density (with respect to $u$), known  as the Bernstein transition function,
   that has been  investigated   in quite divorced from the present study contexts, \cite{jamison,zambrini,gar3}:
 $B(x,t;u,r;y,s) =   k(x,t|u,r)\, k(u,r|y,s)/k(x,t|y,s)$,   where $\int _{\cal{D}}  B(x,t;u,r;y,s)du =1$

\subsection{Conditioning via Bernstein transition function:  dynamical law that underlies  long survival.}

   Let us    adjust the previous notation for the  Bernstein function  for the diffusion process (1)   to refer to an overall
    time interval [0,T], whose  duration  $T$ is arbitrarily assigned,  but assumed  a priori to be large.
     We focus attention on the  transitional $(x,t)$    behavior of the Bernstein function,   for   times      $[s,t] \subset [0,T]$,
      $T\gg t>s\geq 0$),   provided we fix   two   control space-time  points:  $(y,s)$  for the origin,    $(u,T)$ for the target, leaving
            $(x,t)$ as the unrestricted,  "running"  one.

Accordingly, we rewrite the  Bernstein transition function  as follows:
\be
 B(u,T;x,t;y,s) =  \frac{ k(u,T|x,t)\, k(x,t|y,s)}{k(u,T|y,s)} .
 \ee
remebering that  presently  it    is a  probability density  with respect to  $x\in [-c,c]$ (e.g. integrates to $1$).

     Although Eq. (5)  explicitly determines     the  time  $t$  evolution of the   Bernstein density, given $(y,s) and $(u,T),   the main goal
     of our subsequent analysis is to deduce  the detailed (as yet unspecified)
      dynamical rule for the Bernstein density, as a function of $(x,t)$  which would have the form of a standard transport equation,
      appropriate for diffusion processes,
       like e.g. the Fokker-Planck one.

 We point out that an analogous problem  has been addressed in another context, \cite{jamison,zambrini}. The outcome was
the so-called  the Bernstein stochastic process, whose Markovianess could  be  established
 under suitable  (supplementary)  conditions.   \\

Since the time interval $(T-s) \gg s$ is large (and likewise $T-t$) if compared to $T$, the transition density   $k(u,T|y,s)$  stands for the mass of all
   sample paths that have a large  overall survival time $\sim T$, while running from $y$ to $u$ within $(-c,c)$.
    Moreover, the kernel $k(u,T|x,t)$, where $x \in (-c,c)$ may be chosen arbitrarily,
    refers to paths with a large  survival time $\sim T$ as well.  To the contrary, the kernel  $k(x,t|y,s)$  quantifies the mass of sample paths
   that run from $y$ to $x$ in a relatively short (compared to $T$) time interval $t-s$. \\

  Our further  reasoning  relies  on asymptotic (large  time) properties of Feynman-Kac kernels.  For sufficiently large value
  of $T$, the dominant terms in the numerator and denumerator of Eq. (5)  have a similar  form.  An exemplary  asymptotic of
  of the transition density  $k(u,T|y,s)$) reads:
\be k(u,T|x,t) \sim  \sin[{\frac{\pi} L}  (x+c)]\, \sin[{\frac{\pi } L} (u+c)] \, \exp
\left(- {\frac{\pi ^2}{2L^2}} (T-t)\right),  \ee
and  that for  $k(u,T|y,s)$  is   readily  obtained from $k(u,T|x,t))$ by formal  replacements $x\rightarrow y$, $t \rightarrow s$.

Accordingly (remember about $T\gg t$  and $L=2c$),  we arrive at a conditioned transition density
where the    (familiar in the mathematical literature)  Doob-type conditioning  of  $k(x,t|y,s)$  is   non-trivially modified
   by  an   emergent time-dependent factor
  $\exp \left( +{\frac{\pi ^2}{8c^2}} (t-s)\right) $.  We  actually  have:
\be
B(u,T; x,t; y,s) \sim     p(x,t|y,s) =   k(x,t|y,s)  \, {\frac{\sin[{\frac{\pi} {2c}}  (x+c)]}{\sin[{\frac{\pi} {2c}}
(y+c)]}} \, \exp \left( +{\frac{\pi ^2}{8c^2}} (t-s)\right).
\ee
Note that $\sin[{\frac{\pi} {2c}}  (x+c)] = \cos({\frac{\pi} {2c}} x) $.  We recall  that the entry $k(x,t|y,s)$ in the
above is the transition density of the  original  process with absorption at the interval endpoints, see Eq. (1).

     In the large  $T$ regime both  $T$ and $u$ dependence are   absent in the approximate formula (7). Thus, the target point  $u$
is irrelevant for the description of the dynamical behavior  of the Bernstein function  for times $\tau \in [s,t]$.
 There was even no need to execute literally the  $T  \rightarrow \infty $ limit. Once we have $T\gg t$, an approximation  (7)
 is fully legitimate. Moreover the $T\rightarrow  \infty $ limit actually admits time intervals $[s,t]$ of arbitrary finite duration.

An interesting point is that in Eq. (7),  we have arrived at the well known  transition probability density of the unique  Markovian diffusion process
 that never leaves the  prescribed  interval (its  endpoints are inaccessible from the interior), see e.g. \cite{knight,pinsky}
The transition density  (7) of this  conditioned diffusion process is the sought for dynamical law for the time evolution
of the Bernstein density (under our premises). The corresponding transport equation is well know as well.

  By general principles  we  deduce  \cite{knight,pinsky}   the forward drift of the conditioned diffusion process  in question
    \be
  b(x) =  \nabla \ln \cos ({\frac{\pi}{2c}} x) =
  - {\frac{\pi}{2c}} \tan ({\frac{\pi}{2c}} x)
  \ee
 and the transport equation  in the    Fokker-Planck
 form (partial derivatives are executed with respect to $x$):
 \be
 \partial _t  \rho = {\frac{1}2}  \Delta  \rho - \nabla (b\rho),\ee
 The transition density $p(x,t|y,s)$,  Eq. (7),   is  its solution.  We have also  the  probability transport rule  valid
 for any probability density
  $\rho(x,t) = \int _{-c}^c \rho _) (y) p(x,t|y,0) dy$ with $\rho _0(y)$ considered as the initial data for the  F-P evolution.

\begin{figure}[h]
\begin{center}
\centering
\includegraphics[width=60mm,height=60mm]{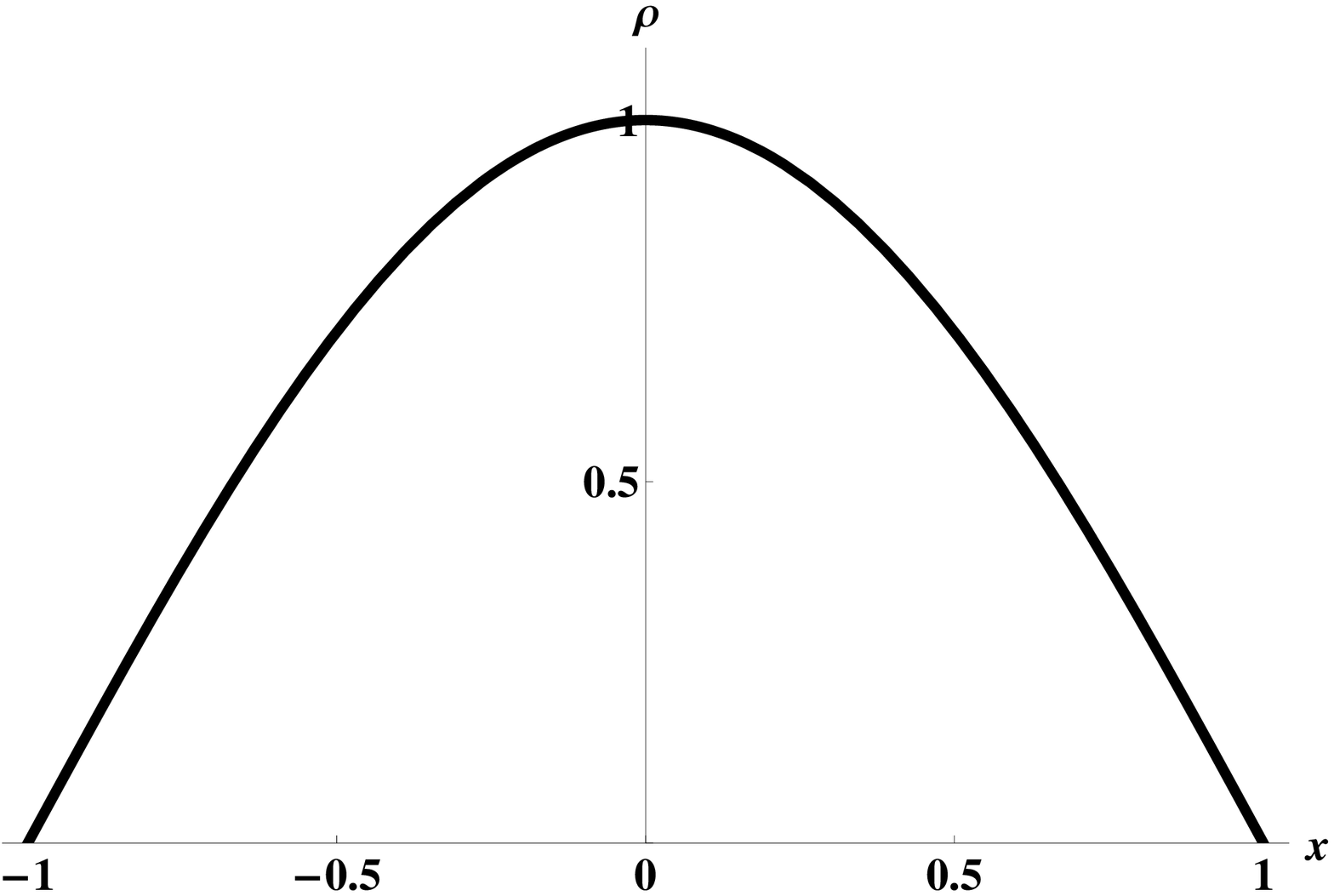}
\includegraphics[width=60mm,height=60mm]{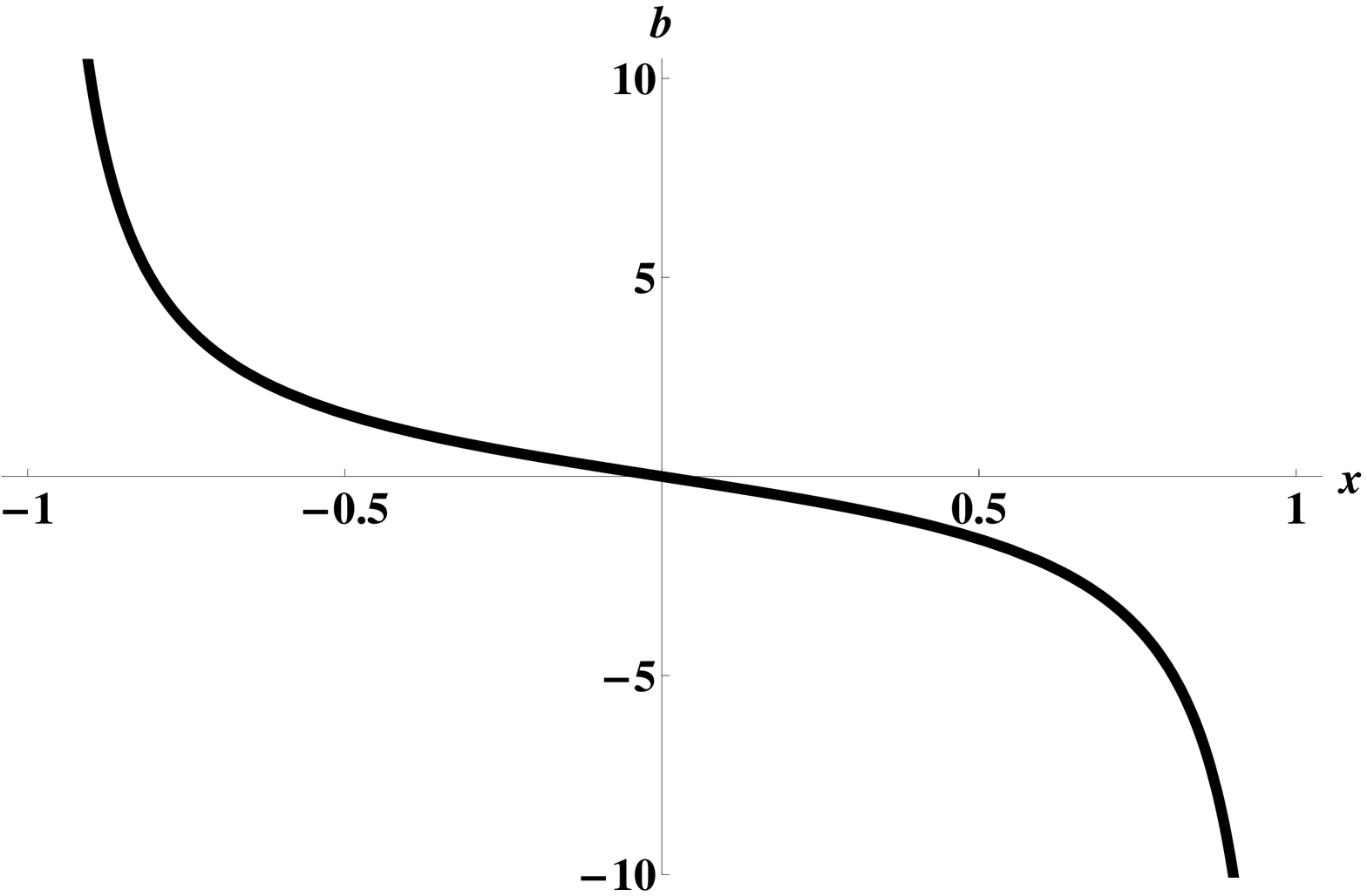}
\caption{Conditioned Brownian equilibrium in the interval with inaccessible endpoints: the permanent trapping enclosure. Left - the probability density
$\rho (x)= cos^2(\pi x/2)$,
right - the forward drift  $b(x) =  - (\pi /2)\,  \tan (\pi x /2) $  of the Fokker-Planck equation  in a trap $(-1,1)\subset R$, \cite{thanks}.}
\end{center}
\end{figure}

 Accordingly, the  dynamics of  the Bernstein function (5), on the $s,t, t-s$ time scales that are small relative to $T$ is   well
  approximated   by that of  a transition probability density (7) of  the  diffusion  process never leaving the interval.
  The  larger is the presumed survival time  $T$   of sample paths in question, the larger may be (ultimately arbitrary, for
   $T\rightarrow \infty $)   the  time  duration of  the   "small scale" $[s,t]$ process.

The  dynamics (7) is thus a generic  (albeit approximate for finite $T$)  property of all long living  trajectories of random motion
 in the interval with absorbing ends. The requirement of an infinite survival time (permanent trapping)  leaves us with the
 diffusion process (7), with a guarantee that   no trajectory may reach the interval endpoints.

\subsection{Feynman-Kac kernel in the interval and eigenfunction expansions.}

We point out an obvious link of the previous analysis  with the standard (quantum mechanical by provenance) spectral problem
for the operator  $ - {\frac{1}2}  \Delta $ in   an infinite well, supported on the interval $(-c,c)$.  Denoting $\lambda _n$, $n=1,2,...$
the eigenvalues and  $\phi _n,  n=1,2,...$  the orthonormal basis system composed of the eigenfunctions $\phi _n$.

In particular  a   "miraculous" emergence of the time-dependent factor  $\exp \left( +{\frac{\pi ^2}{2L^2}} (t-s)\right)$
might seem to be annoying at the first glance. However  this factor secures the existence of an asymptotic   invariant probability density.
Since $\pi ^2/2L^2 = \lambda _1$  is the lowest eigenvalue of $ - {\frac{1}2}  \Delta _{\cal{D}}$, we effectively  encounter here a standard   additive
 "renormalization"  $ - {\frac{1}2}  \Delta _{\cal{D}} - \lambda _1$ in the semigroup generator definition. Then, the  strictly positive operator
  (bottom of the spectrum is $\lambda _1$) becomes replaced by the non-negative operator with the bottom of the spectrum at zero.
   See e.g. \cite{faris,roepstorff}.

The  spectral decomposition of $(1/2) \Delta _{{\cal{D}}}$    allows  to rewrite   $k(y,s|x,t)$   in a handy form:
  \be
  k(x,t|y,s) = \sum _{n=1}^{\infty } e^{-\lambda _n (t-s)}\, \phi _n(x) \phi  _n(y).
  \ee

It is clear that under suitable regularity assumptions concerning the  long time  $T$ behavior of Feynman-Kac kernels,
 specifically that we have  $k(y,s|u,T) \sim \phi _1(y) \phi _1(u) \exp (-\lambda _1(T-s))$, one arrives at  the legitimate
 $T\rightarrow \infty $ expression.
 \be
p(x,t|y,s) = k(y,s|x,t) {\frac{\phi _1(x)}{\phi _1(y)}} e^{+\lambda_1  (t-s)}
\ee
as anticipated previously (set  e.g. $\lambda _1= {\frac{\pi ^2}{2L^2}}$
and $\phi _1(x)= \sqrt{{\frac{2}L}} \,   \cos(\pi x/L)$).\\

Given the limiting transition probability density (11), we can now assume that $t\gg s$, By resorting to the large time behavior of the kernel
 (6), we  readily deduce the relaxation pattern of the random motion. Namely, we get:
    \be  p(x,t|y,s,t)  \sim   (\phi _1(x))^2 = \rho (x)   =  {\frac{2}L} \,   \cos^2(\pi x/L)
 \ee
While keeping in memory  the   $L^2(-c,c)$ normalization of $\phi _1(x)$  we  have thus identified the  stationary invariant
  probability density   $\rho (x)$ of the  diffusion  process, conditioned never to leave the interval. \\

\subsection{Time  rates to equilibrium.}

Collecting together the generic features (typicalities)  of  the  diffusion in the interval with absorbing ends,  i.e.
  (i)  $k(x,t|y,0)= \sum_j \exp(- \lambda  _j t) \, \phi _j(y) \phi _j(x)$,  (ii)  the  transition density of the conditioned process  in  the universal form
    $p(x,t|y,0) = k(x,t|y,0) \, e^{+\lambda_1  t}  \,   \phi _1(x) /\phi _1(y) $ and  (iii) the  relaxation  asymptotic
     $\lim _{t\rightarrow \infty }p(x,t|y,0) = \rho (x)= \phi _1^2(x)$, we can address an issue of the  time
    rates towards equilibrium in the conditioned random motion   (11).

 Let us denote $\tilde{k}(x,t|y,0) = e^{\lambda _1t} k(x,t|y,0)/ \phi _1(x) \phi _1(y)$.
 We employ an estimate (known to be valid for $t>1$), \cite{smits}, see also \cite{lorinczi}:
\be
|\tilde{k}(x,t|y,0) -1| \leq C\, e^{-(\lambda _2- \lambda _1)t}
\ee
where $C$ is a  suitable constant.
That yields immediately (multiply both sides by $\rho (x)= \phi _1^2(x)$) the general formula for
 the time rate to equilibrium, provided we name  the invariant stationary density  $\rho (x)$   the equilibrium
  density of the  process:

\be
 |p(x,t|y,0) - \rho (x)| \leq    C\, e^{-(\lambda _2- \lambda _1)t}  \, \rho (x)  .
 \ee
We have the exponentially fast approach toward equilibrium  (realized via the  invariant density
shape rescaling by en exponential factor), whose speed depends on the spectral gap between two lowest eignvalues.

In the interval $(-1,1)$ we have  (compare e.g. Eq. (1))  $\lambda _1= \pi ^2/8$  and $\lambda _2 = \pi ^2/2$,
hence $\lambda _2 - \lambda _1 = 3\pi ^2/8$. We have also    $\rho (x)= \cos ^2(\pi x/2)$.

The above rate formula has much broader significance, since  it is generally valid for diffusions in convex domains
(irrespective of space  dimensions), \cite{smits}.\\

\subsection{Semigroup transcript of the Fokker-Planck dynamics.}

Let us employ the standard text-book notation \cite{risken}  with dimensional constants kept explicit.
Scalings  leading to  dimensionless notations,   used throughout the present paper, are obvious.

Let us  consider the Langevin equation  $dX_t = b(X_t) dt + \sqrt{2\nu } B_t$, where the drift
 velocity is a gradient field, e.g. $b= - \nabla {\cal{V}} $. The  correspoding  Fokker-Planck equation takes the
 form  $\partial _t \rho = \nu \Delta \rho - \nabla (b\rho )$.

  We assume that asymptotically, the  Fokker-Planck  dynamics sets down at the equilibrium (stationary)
   solution $\rho _*(x)$, e.g. $\lim _{t\rightarrow  \infty } \rho  (x,t) = \rho _*(x)$.   Since the drift $b(x)$ does
    not depend on time, the Fokker-Planck equation implies that  $b= \nu \nabla \ln \rho _*$, i.e. the
     stationary solution  $\rho _*$  fixes $b$ and in reverse. Accordingly, we arrive at the  Boltzmann-Gibbs form of
      $\rho _* (x) = \exp [- {\cal{V}}(x)/\nu ]$.  That  in consistency with   the primary definition of
       $b=  - \nabla {\cal{V}}$.

 Following tradition \cite{risken}   let us introduce a multiplicative decomposition of $\rho (x,t)$:
 \be
 \rho (x,t) = \Phi (x,t) \rho _*^{1/2}(x).
\ee
So introduced positive function  $\Phi (x,t)$ satisfies the generalized diffusion equation:
\be
\partial _t \Phi = \nu \Delta \Phi -  V \Phi ,
\ee
where the potential field  $V=V(x)$   is given  (up to an additive constant allowing to  make positive any bounded from below $V$):
\be
V = {\frac{1}2} \left( {\frac{b^2}{2\nu }} + \nabla b\right) .
\ee

Eq. (15) actually derives form  the semigroup dynamics $\exp (-t\hat{H} /\nu )$  describing a stochastic process with killing, if
 $V(x)$ is positive-valued (that is trivial for potentials which are bounded from below, since we can  always
add a suitable constant to make the potential positive-definite).

If the spectral solution for $\hat{h}=- \nu \Delta  + V$   allows for  an isolated eigenvalue at the bottom of
 the spectrum, denote it $\lambda _1$, we can always introduce $\hat{H} - \lambda _1$  as the semigroup generator.
 Then (preserving only the discrete part of the spectral decomposition):
 $\Phi (x,t) = \exp (+\lambda t) \sum _{n=1} c_n  \exp(-\lambda _nt)\, \phi _n(x) \rightarrow \phi _1(x) = \rho _*^{1/2}(x)$
 In that case the dynamics of $\Phi(x,t)$ asymptotically sets down at  $\rho _*^{1/2}(x)$. \\

{\bf Remark:} It is an amusing exercise to check that by setting $\nu =1/2$ and inserting the drift field expression
(8), we actually obtain $V(x)=0$ identically in the open interval $(-c,c)$.  Note also that $b = \nabla \ln  \rho _*$, where
$\rho _* =  (\phi _1)^2$, see also  \cite{gar1}.

\section{Disk with an absorbing boundary and the conditioning.}

  To make clear the links with the  past and current   mathematical literature on a similar subject,
 let us e.g. quote the major result of the paper \cite{banuelos},
 see also \cite{banuelos1}  and \cite{lorinczi0,lorinczi},   which has been actually formulated for bounded planar domains.

We adopt the original  notation of Ref. \cite{banuelos}   to that used in the present paper and stress that the order of
variables $x$ and $y$
is  here  interchanged if compared with that in  Ref. \cite{banuelos}, formula (1.1).  Namely, let  $\Omega $  stand for a
planar domain of finite area, $\lambda $ is the first positive eigenvalue of half the Laplacian $(1/2)\Delta $ in $\Omega $,
and  $\phi $ is the first $L^2$ normalized eigenfunction $\int _{\Omega } \phi ^2=1$.

 Let $k(x,t|y,o)$ be the fundamental solution
of the heat equation with Dirichlet boundary conditions. Then for any $x\in \Omega $, there holds:
 \be
\lim _{t\rightarrow \infty } {\frac{e^{\lambda t}\, k(x,t|y,0)}{\phi (y) \phi (x)}}=1
\ee
uniformly in $y \in \Omega $.   Eq. (13) clearly  is  instrumental in defining the Brownian motion conditioned to stay forever
 in $\Omega  $.  Let us recall then long time behavior of the transition kernel as reported in eq. (7).
  The same asymptotic  in the present notation would read:
 $k(x,t|y,0) \sim \phi (y) \phi (x) \,  \exp (-\lambda t)$.
 \\

\begin{figure}[h]
\begin{center}
\centering
\includegraphics[width=60mm,height=60mm]{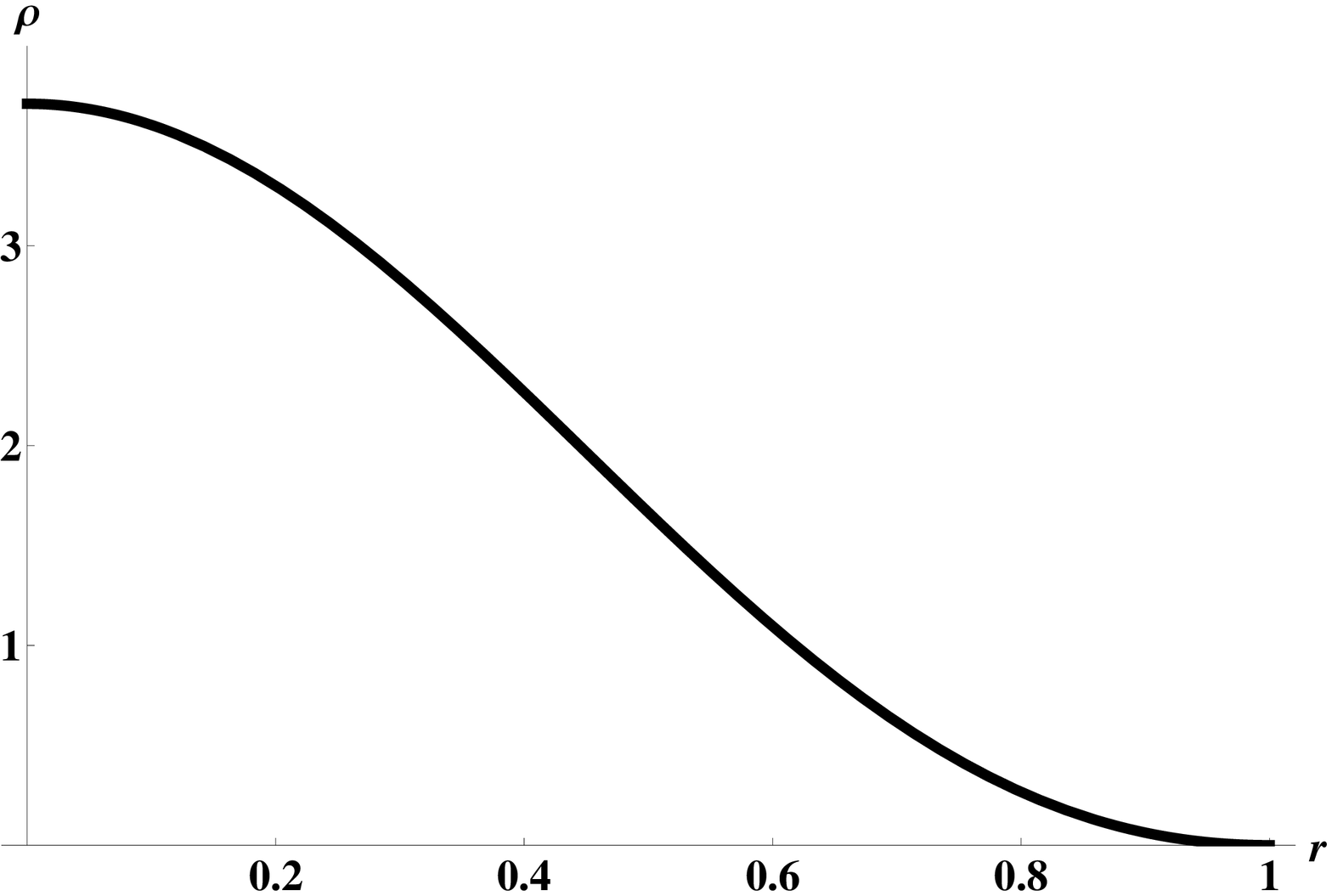}
\includegraphics[width=60mm,height=60mm]{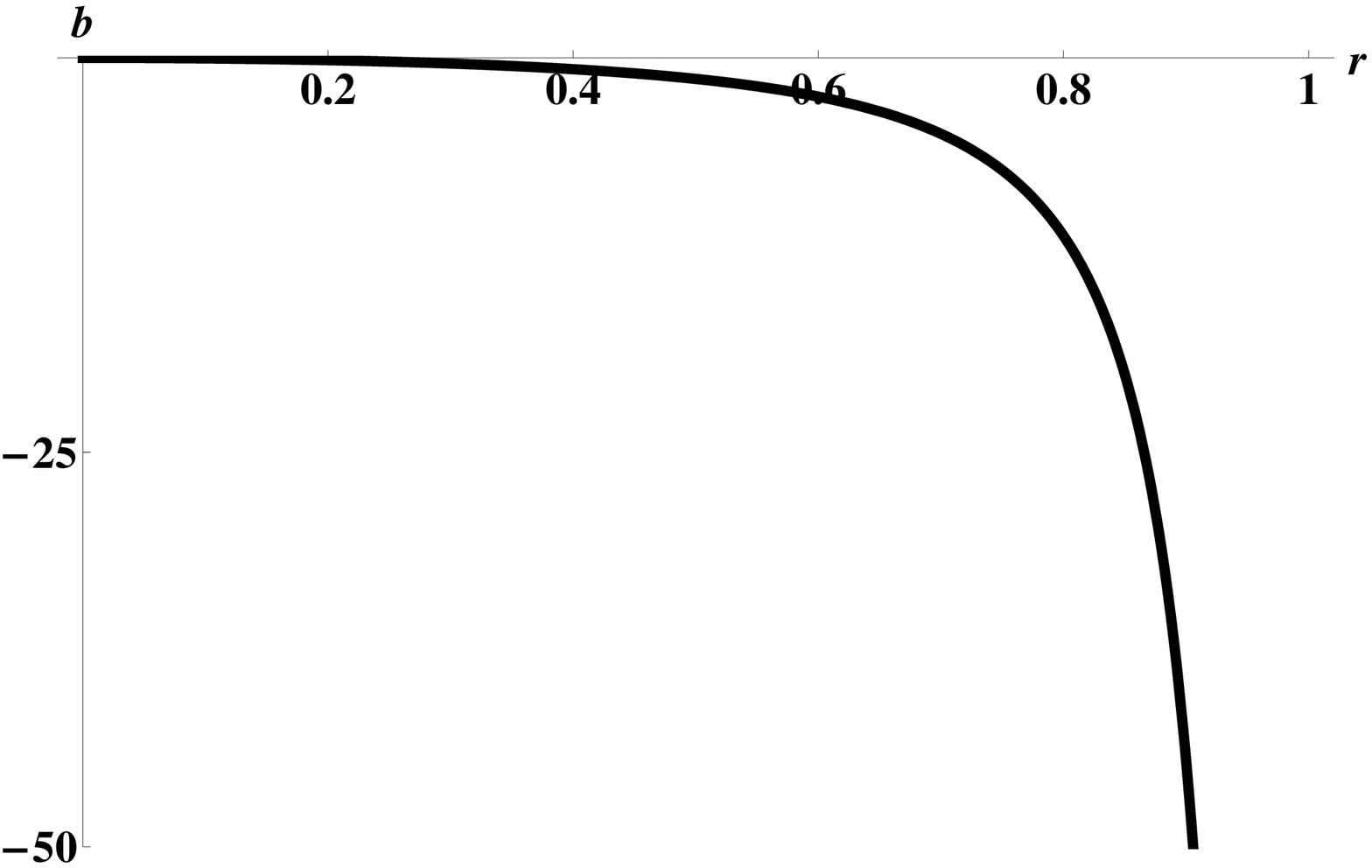}
\caption{Conditioned Brownian equilibrium in  the unit  disk  with an  inaccessible  boundary: the permanent trapping enclosure.
 Left - the probability density $\rho (x)$,
right - the forward drift  $b(x)$  of the Fokker-Planck equation  in a trapping disk, \cite{thanks}.}
\end{center}
\end{figure}

As a simple example of the bounded (and convex) planar domain,  we take  a domain in the regular disc shape, i.e. bounded by the
 circle of a fixed radius $R$.   The spectral solution for such $2D$   domain is  clearly in the reach,   albeit somewhat murky
   from the   casual  (user  friendly)  point of view.\\
The  spectral decomposition  formula (8)   for the transition density of the process  in a disk  with absorbing boundary  is here valid,
with suitable  amendments that  reflect the two-dimensional setting.

Since we are interested in the long $T$ duration  of the conditioned process,  in view of our previous discussion in Section II,
  we need basically the knowledge
of the  stationary density   and the forward drift. In the present case these read  respectively:
\be
\rho ({\bf r}) \sim {\frac{j_0^2\left(\frac{z_1 r}{R}\right)}{j_1^2(z_1)}}  =  \phi _1^2({\bf r}),
\ee
where $z_1=2.4048...$  is the first positive zero of the Bessel function $j_0(r)$, and  ${\bf {r}}= (x,y)$
 (in below we shall use  $\hat{\bf r} = (1/r) (x,y)$). The density is purely radial, hence
 only the radial component of ${\bf  b}({\bf r})= b({\bf r})\, \hat{\bf r} $  is different from zero with
 \be
 b({\bf r}) = {\frac{\partial }{\partial r}}\,   \ln \phi _1({\bf r}) =  -  {\frac{ z_1 j_1\left(\frac{z_1 r}{R}\right)}{R j_0(z_1r)}}.
 \ee
We have here a fully-fledged two-dimensional Brownian motion,  with a drift  ${\bf b({\bf r})}$  that is a purely radial vector
 field. The Fokker-Planck equation is two-dimensional:  $\partial _t \rho ({\bf r},t) =  (1/2) \Delta \rho - {\bf  \nabla } ({\bf b} \rho )$.
 For completness, we shall reproduce this equation in the polar  coordinates, with the  radial form of the drift being explicit:
 \be
 \partial _t \rho = \frac{1}{2}   \left( {\frac{\partial ^2}{\partial  r^2}} + {\frac{1}{r}}  {\frac{\partial }{\partial r}} \right)  \, \rho
- \left( {\frac{1}r} {\frac{\partial (r  b \rho ) }{ \partial _r}} \right)
\ee
We point out that the time rate formulas become more complicated in dimensions exceeding one  $D>1$, since  if we admit the  general
(not exclusively
 radial)  random motion  the  disk, the Laplacian   spectrum becomes degenerate (except for the ground state) and the
 eigenvalues  (increasing)  order  is not set merely by $n=1,2,...$, but
  needs to account for the angular label $l$, \cite{acqua,bickel,robinett}.   E.g. we know the the first positive zero of $j_0$  ($l=0$ sector)
   approximately    equals $2.4048$, the second  zero equals $5.5201$, while the first zero of $j_1$  (l=1 sector)  reads $3.8137$,
    and the first zero of $j_2$ equals $ 5.1356$.

  {\bf Remark:}  The  survival problem  on the  disk  can be reformulated in a pictorial way by referring directly to the front-cover picture
   of Redner's monograph \cite{redner} and taking some inspiration from \cite{redner1, redner2}, even though the problems addressed
   there refer to absorbing boundary conditions and first passage time issues.
   Take literally a (somewhat drunken)   Brownian wanderer on top of the island  (disk)   which  is  surrounded by a  cliff
   (plus predators in the ocean).   Add a psychologically motivated fear component to the wanderer behavior,  when one is close
    to the  island  boundary.  Then we readily arrive at the following problem:
   how should that fearful Brownian agent move to increase his chances for survival. Our  disk solution, establishing the regime
    of the inaccessible boundary,      gives  the  optimal  stationary probability danesity
   (transition density likewise, albeit not in a handy  closed  form)  and the drift field.   The  random  wanderer  obeying the associated
    Fokker-Planck equation and following the  underlying  Langevin)  dynamics  would  live indefinitely on the island (disk), while   moving  safely away from the inaccessible
    boundaries with no chance to  reach them.

As a  useful supplement, let us add that the $3D$ spherical well
problem, with absorption at the boundary,\\
  can be addressed in the very same way. The probability density reads  (here ${\bf r}= (x,y,z)$  and $\hat{\bf r}$ stand s for a unit vector):

\be  \rho ({\bf r}) = {\frac{1}{2\pi R}}\, {\frac{\sin ^2 (\pi
r/R)}{r^2}} = \phi _1^2({\bf r})
\ee

where $R$ stands for the sphere radius.  The forward drift of the
pertinent stochastic process, that never leaves the sphere interior,
reads:

\be {\bf b}({\bf r}) =   {\frac{\partial _r
\phi _1 ({\bf {r}})}{\phi _1 ({\bf {r}})}}\, {\hat{\bf r}} =   \left[
{\frac{\pi}{R}} \cot \left({\frac{\pi r}{R}}\right)  - {\frac{1}r}\right] {\hat{\bf r}}  = b({\bf r}) {\hat{\bf r}}
\ee
We have here a fully-fledged three-dimensional Brownian motion,  with a drift  ${\bf b({\bf r})}$.
The Fokker-Planck equation is now three-dimensional and its radial form is reproduced for completness:
 \be
 \partial _t \rho = \frac{1}{2}   \left( {\frac{\partial ^2}{\partial  r^2}} + {\frac{2}{r}}  {\frac{\partial }{\partial r}} \right)  \, \rho
- \left( {\frac{1}r^2} {\frac{\partial (r^2  b \rho ) }{ \partial _r}} \right).
\ee

\section{Feynman-Kac kernels for killed stochastic processes and the killing removal.}

It is well known that operators of the form $\hat{H}= - (1/2)\Delta + V  \geq 0$  with $V\geq 0$  give rise to transition kernels of diffusion -type  Markovian processes with killing (absorption), whose rate is determined
 by the value of $V(x)$ at $x\in R$. That interpretation stems from the celebrated Feynman-Kac (path integration) formula, which assigns to $\exp(-\hat{H}t)$
 the positive  integral kernel.

   \be  [\exp(- (t-s)(-{\frac{1}2}\Delta  + V)](y,x)  =
       \int   \exp[-\int_s^t V(\omega(\tau  )) d\tau ]\,   d\mu _{s,y,x,t}(\omega)
        \ee
 In terms of Wiener paths that kernel is constructed as a path integral over paths that get killed at a point $X_t=x$
 with an extinction  probability $V(x)dt$,    in the time interval$(t,t+dt)$ (note that physical  dimensions of of $V$ before
  scaling them out, were $J/s$, that is usually secured by a factor $1/2mD$  or $1/\hbar$).
   The killed path is henceforth removed from the ensemble of on-going Wiener paths.

Given a discrete  spectral solution for $\hat{H}= - (1/2)\Delta + V$   with $V(x)  \geq 0$, comprising  the monotonically
 growing series of non-degenerate positive eignevalues, with real  $L^2(R)$ eigenfunctions.  The  integral kernel of $\exp(-t\hat{H})$ has the  time-homogeneous form
\be
k(y,x,t)= k(x,y,t)= \sum_j \exp(- \epsilon _j t) \, \phi _j(y) \phi _j(x).
\ee

Consider  the harmonic oscillator problem with  $\hat{H}= (1/2)(-\Delta + x^2)$.
The integral kernel of $\exp(-t\hat{H})$  is given by a  classic   Mehler formula:

\be k(x,y,t) = [\exp(-t\hat{H})(y,x)=   {\frac{1}{\sqrt{\pi }}}
\exp[- (x^2+y^2)/2] \, \sum _{n=0}^{\infty } {\frac{1}{2^n n!}}
H_n(y) H_n(x) \exp(-\epsilon _n\, t) = \ee
$$
 \exp(-t/2)\, (\pi [1-\exp(-2t)])^{-1/2} \exp \left[{\frac{1}2} (x^2-y^2) -
   {\frac{(x- e^{- t}y)^2}{(1- e^{-2 t})}}\right]  =
 $$
$$
 {\frac{1}{(2\pi  \sinh t)^{1/2}}}  \exp  \left[ - {\frac{(x^2+y^2)\cosh t  - 2xy}{2\sinh t }} \right] ,
 $$

where $\epsilon _n= n +  {\frac{1}2}$,  $\phi _n(x) = [4^n (n!)^2 \pi ]^{-1/4} \exp(-x^2/2)\, H_n(x)$ is the $L^2(R)$
 normalized Hermite  (eigen)function, while $H_n(x)$ is the n-th Hermite polynomial
  $H_n(x) =(-1)^n (\exp x^2) \, {\frac{d^n}{dx^n}} \exp(-x^2) $.  Note a conspicuous presence of the time-dependent factor $\exp (-t/2)$.

Let us replace $t$ by   $T-t$ and accordingly consider
$k(x,T|y,t)=k(T-t,x,y)$. Now we pass to the conditional transition
density

\be B(u,T;x,t;y,s)= {\frac{k(y,s|x,t) k(x,t|u,T)}{k(y,s|u,T)}} \ee
  and investigate the   large $T$ (eventually $T \rightarrow  \infty $) regime. Since for large  value $T$ we have  (comapare e.g.  also
   \cite{roepstorff})
\be k(x,t|u,T) \sim   {\frac{1}{\sqrt{\pi }   e^{(T-t)/2}}}\,
e^{-{\frac{1}2} (x^2+ u^2)}, \ee by repeating the conditioning
procedure of Section II, we readily arrive at the approximation
(c.f. (7)) of the Bernstein transition fuction by a
  transition probability density $p(x,t|y,s) = p_{t-s}(x|y)$  of the  familiar  Ornstein-Uhlenbeck process

\be
 B(u,T;x,t;y,s) \rightarrow p(x,t|y,s) =  k(x,t,y,s) \,  \frac{\exp(-x^2/2)}{\exp{(-y^2/2)}}\,
  e^{(t-s)/2} = \ee
$$
k(y,x,t)\, {\frac{\phi _1(x)}{ \phi _1(y)}}\,  e^{+\epsilon _1 (t-s)} =
 [\pi  (1- e^{-2 (t-s)})]^{-1/2}\,
  \exp \left[ -  {\frac{(x- e^{- (t-s)}y)^2}{(1- e^{-2( t-s)})}}\right] .
   $$
where  $\phi _1(x)=\pi ^{-1/2} \exp(-x^2/2)$ and  $\epsilon _1 = 1/2$ have been accounted for.

Here the Fokker=Planck operator takes the form $L_{FP} = (1/2)\Delta -
\nabla   [b(x) \, \cdot ]$   and $b(x)= -x$.
 Clearly $b(x)= \nabla \ln \phi_1(x) $, as should be.
 The asymptotic (invariant, stationary) probability density of the
 pertinent process reads
$ \rho (x)= \phi _1^2(x)=(1/\pi )\,
 \exp(-x^2)$.
For earlier  considerations on how to transform the Feynman-Kac averages for processes with killing
into those that refer to processes without any killing see e.g. \cite{klauder}.

\section{Handling the sink: survival on the half-line and   Bessel processes.}

The   conditioning procedure up to now seems to rely heavily on  contracting   semigroups, whose generators have purely discrete spectral solutions
  with  the bottom eigenvalue being well  separated from the rest of the spectrum.

 The   above outlined conditioning  procedure will surely work for   potentials that are bounded from below continuous fucntions, since we  can
   always redefine   potentials with a bounded from below  negative part,
   by adding to $V(x)$ a modulus of its minimal value  $|V(x_{\min})|$ or a modulus  of any of  its multiple (identical) local minima:
  $V(x) \rightarrow V(x) + |V(x_{\min})|$ so arriving at $V(x) \geq 0$.
 The redefined potential  is nonnegative  and gives  rise to the diffusion-type process with killing, whose transition density  $k(y,s|x,t)$
    is given by the Feynman-Kac formula, \cite{faris,roepstorff}.

  It is our aim to demonstrate that we need not to have  a discrete spectral solution at hand.  The  employed   conditioning
     procedure  appears to work properly,  also when  the  spectrum of the involved  semigroup generator is continuous.
     This  is  e.g.  the case   for   Brownian motion on a half-line with   an  absorbing barrier (sink).

We set the sink at $0$  and consider the Brownian motion as being
restricted to the positive semi-axis ($x\in R^+$).
The pertinent transition density is obtained via the method of
images, by employing the  standard Brownian transition probability density  (induced by $(1/2)\Delta $)
  \be
  p(x,t|y,s) = [2\pi (t-s))]^{-1/2} \, \exp [ - (x-y)^2/2\pi (t-s)]  .
   \ee
  Namely:
  \be
   k(x,t|y,s)=   p(x.t|y,s) -  p(x,t|-y,s)  = {\frac{2}{\sqrt{2\pi (t-s)}}} \,
   \exp [-{\frac{x^2 +y^2}{2(t-s)}}]\,  \sinh  \left( {\frac{xy}{t-s}} \right ) .
  \ee
The large $T$ behavior of $k(u,T|y,s)$ is easily inferred to imply  (the situation is less straightforward than
in the previous examples, since the time label $T$ persists in the exponent), note that we replace $T-s$ and $T-t$ by $T$):

\be
k(u,T|y,s) \sim {\frac{2}{\sqrt{2\pi T}}} \exp[- {\frac{y^2 +
u^2}{2T}}]\, \,  {\frac{u y}{T}}  .
     \ee
  Accordingly,   in the large $T$ regime, the Bernstein function is approximated by an  intriguing functional expression:
\be B(u,T;x,t;y,s) \sim p(x,t|y,s)  \exp [{\frac{y^2-x^2}{2T}}] \ee
where  we single out,  as a leading factor, an immediately
recognizable transition probability function   of the Bessel process
\be p(x,t|y,s) = k(x,t|y,s) \,{\frac{x}y} . \ee

      We have fixed the $(y,s)$ and $(u,T)$ data  and  there is no $u$-dependence in the asymptotic expression  (34). Consequently,
 $\exp (y^2/2T)$  is irrelevant as far as the large $T$ regime is concerned.  What however matters  is the remaining term $\exp(-x^2/2T)$
 which is not quite innocent for $x$'s that are comparable in size with $T^{1/2} $.  Surely, for    large finite $T$ the exponential term may be
 regarded to  very be close to one for   not too large values of $x$, say $x<  T^{1/2}/100$, since $ \exp (-1/20000)\sim 1- 5\cdot  10^{-5}$
 is sufficiently close to $1$.

 Pushing  $T   \rightarrow \infty $ refers to an eternal survival and involves the   point-wise convergence of  $p(u,T;x,t;y,s)$  to $p(x,t|y,s)$  in
Eq. (34),  ultimately  leaving us with a transition probability density for the Bessel process.

  The forward drift of this process is   known \cite{knight,pinsky}  to be equal
   $b(x)= \nabla  \ln x  = 1/x$. The  Fokker-Planck  generator takes the familiar (Bessel process) form $L_{FP} =
(1/2)\Delta - [b(x) \cdot ]$.   We note that the point $0$ is  presently inaccessible for the
  process.

Told otherwise: the one-dimensional Brownian motion starting from $y>0$, conditioned to remain positive  up to time $T$, converges as $T\rightarrow \infty $
  to the  radial process of the three-dimensional Brownian motion, known as the
  Bessel  process.\\

{\bf Remark:}   In the one-parameter family  of Bessel processes, with  drifts of
  the form $b(x)= (1+2a)/2x$,
 in case of   $a\geq 0$, the point $x=0$ is never reached from any $y>0$   with probability
  one.
  To the contrary, for   $a<0$, the barrier at $x=0$  is absorbing   (sink).

  Let us recall the backward generator of the process: $(1/2) \Delta + b(x) \nabla $ with $b(x) = (1+2a)/2x$.
  The  one-parameter family of  pertinent  transition densities reads:
  \be
  k_a(y,s|x,t)= {\frac{y^{2a+1}}{t (xy)^a}}\,  \exp \left[-{\frac{x^2+y^2}{2t}}\right]\, I_{|a|}
   \left({\frac{xy}{2t}}\right) .
\ee
Let us consider the special case of $a=\pm (1/2)$ for which the modified Bessel function takes  a handy form
\be
 I_{1/2}(z) = \sqrt{\frac{2}{\pi z}}\, \,  \sinh z  .
\ee
It is easy to verify that  $k_{-1/2}(y,s|x,t)$   coincides with the transition density of the Brownian
 motion  constrained to stay on   $R+$, with a sink at $0$.   The generator simply is  $( 1/2)\Delta $   on $R^+$, with  absorbing boundary at $0$.

 On the other hand $k_{+1/2}(y,s|x,t)$   is a transition probability density of the
  Bessel process   with $b(x)= 1/x$  (e.g. the Brownian motion   conditioned to never reach $0$, if started from any $y>0$).
    Its F-P generator reads $(1/2)\Delta  - \nabla (b(x)\cdot ) $,      with $b(x)=\nabla \ln x$.

\subsection{Prospects}

Our  conditioning strategy involving the Bernstein transition probability densities heavily relies on the large time asymptotic properties
of transition densities for processes with  absorption (killing). Once the kernels $k(y,s|x,t)$ are given the spectral resolution form
 (e.g. the eigenfunction expansions) we can expect that  our considerations should be safely extended  to  Levy-stable  processes,
  additively  perturbed  by suitable potential functions.
That refers to L\'{e}vy motions in energy landscapes of ref. \cite{lorinczi,lorinczi0}  and references therein.

In case of a discrete spectral  resolution of the random motion generator, transiton kernels are in principle computable and amenable
to asymptotic procedures of Sections II to V.   Basically, the  kernels have no  known explicit  analytic forms.
If we now the lowest eigenvalue and the corresponding eigenfunction, the conditioning itself can be imposed in exactly the same way as before
 (via Bernstein transition functions).
Link of the killed  stochastic process  and its "eternally living" partner  can surely be established  for jump-type processes as well.

For deceivingly simple problems   (albeit in reality,  technically quite involved) of the L\'{v}y-stable
 finite and  infinite well or its spherical well analog, numerically accurate and  approximate analytic
  formulas  are known for the ground states. We have in  hands their shapes (hence the resultant stationary
  probability densities of the conditioned process)  together with corresponding lowest eigenvalues,  \cite{zaba,zaba1,zaba2,zaba3}.
  Analogous results were established for some L\'{e}vy stable oscillators, \cite{stef1,lorinczi1,lorinczi2,zaba}

   The solution for the half-line  L\'{e}vy-stable  problem with absorbing barrier (rather involved
  and available in terms of an  approximate analytic expression) is also in existence,   \cite{stos},
    and may  be used to deduce the process living eternally on the half-line, following our conditioning method.
  All that needs more elaborate analysis which we relegate to the future.


\begin{thebibliography}{99}
\bibitem{redner}  S. Redner, {\it A Guide to First-passage Processes}, (Cambridge University Press,
Cambridge 2001).
\bibitem{redner1} S. Redner, P. L. Krapivsky,  "Diffusive escape in a nonlinear shear flow. Life and death  at the
 edge of a  windy cliff", J. Stat. Phys. {\bf 82}, 999, (1996)
\bibitem{redner2} P.L. Krapivsky, S. Redner, "Life and death in an expanding  cage  and at the edge of  a receding
 cliff", Am. J. Phys. {\bf 64}, 546,  (1996)
\bibitem{agranov} T. Agranov, B. Meerson and A. Vilenkin, "Survival of interacting diffusiong particles inside a domain with absorbing boundary",
Phys. Rev. B {\bf 93}, 012136, (2016).
\bibitem{redner0} B. Meerson and S. Redner, "Large fluctuations in diffusion-controlled absorption", J. Stat Mech. (2014), P08008,
\bibitem{pekalski} M. Droz and A. Pekalski, "Fast power law-like decay for a diffusive system with
absorbing borders", Physica A {\bf 470}, 82, (2017).
\bibitem{li}  M. Li, N. D. Pearson, A. M. Poteshman, "Conditional estimation of diffusion processes", J. Financial Economics {\bf 74},  31-66,
(2004).
\bibitem{castro} L. B. Castro adn A. S. de Castro, "Trapping of a particle in a short-range harmonic potential well", J. Math. Chem, {\bf 51}, 265, (2013).
\bibitem{lukic} B. Luki\'{c} et al., "Motion of a colloidal particle in an optical trap", Phys. Revv. E {\bf 76}, 011112, (2007).
\bibitem{gar} Ph. Blanchard and  P. Garbaczewski, "Natural boundaries for the Smoluchowski equation and affiliated diffusion processes",
    Phys. Rev. E {\bf 49},  3815, (1994).
\bibitem{gar1} P. Garbaczewski and M. \.{Z}aba, "Nonlocal random motions and the trapping problem",  Acta Phys. Pol. B {\bf 46}, 231, (2015).
\bibitem{vilela} S. Eleuterio and  R. Vilela Mendes,  "Reconstruction of the dynamics from an eigenstate"  J. Math. Phys. {\bf 27},178, (1986)
\bibitem{vilela1}  S. Eleuterio and  R. Vilela Mendes,  "Stochastic ground state processes",    Phys. Rev. {\bf B 50}, 5035, (1994)
\bibitem{roepstorff}  G. Roepstorff, {\it Path integral approach in quantum physics}, (Springer-Verlag, Berlin, 1994).
\bibitem{faris} W.G. Faris, "Diffusive Motion and Where It Leads", in: W.G. Faris (Ed.),
{\it Diffusion, Quantum Theory and Radically Elementary Mathematics},
(Princeton University Press, Princeton 2006).
\bibitem{davies} E.B. Davies, {\it Heat Kernels and Spectral Theory}, (Cambridge University Press,
Cambridge 1990).
\bibitem{knight}  F. B. Knight, "Brownian local times and taboo processes", Trans. Amer. Math. Soc., {\bf 143}, 173, (1969).
\bibitem{pinsky} R. G.Pinsky, "On the convergence of diffusion processes conditioned to remain in a bounded region for large time to limiting
positive recurrent processes", Annals of Probability, {\bf 13}(2), 363, (1985).
\bibitem{korzeniowski} A. Korzeniowski, "On diffusions that cannot escape from a  convex set", Statistics \&  Probability Letters, {\bf 8}, 229, (1989).
\bibitem{jamison}  B. Jamison, "The Markov processes of Schr\"{o}dinger", Z. Wahrsch. verw. Gebiete, {\bf 32}, 323, (1975).
\bibitem{risken} H. Risken, {\it The Fokker-Planck Equation}, (Springer-Verlag, Berlin, 1989)
\bibitem{zambrini}  J. C. Zambrini,  "Stochastic mechanics according to E. Schr\"{o}dinger",  Phys. Rev. A {\bf 33}, 1532, (1986).
\bibitem{gar3} P. Garbaczewski and J-P. Vigier, "Brownian motion and its descendants according to Schr\"{o}dinger", Phys. Lett. A {167}, 445, (1992).
\bibitem{thanks}  Courtesy of Dr M. \.{Z}aba.
\bibitem{smits} R. G. Smits, "Spectral gap and rates to equiibrium for diffusions in convex domains", Michigan Math. J., {\bf 43}, 141, (1995)
\bibitem{banuelos} R. Ba\~{n}uelos and B. Davis, "Heat kernel, eigenfunctions and conditioned Brownian motion in planar domains",
J. Funct. Anal. {\bf 84}, 188, (1989)
\bibitem{banuelos1}   R. Ba\~{n}uelos, "Intrinsic ultracontractivity and eigenfunction estimates for Schr\"{o}dinger operators",
 J. Funct. Anal. {\bf 100}, 181, (1991)
\bibitem{lorinczi0} K. Kaleta and J. L\H{o}rinczi, "Pointwise eigenfunction estimates and intrinsic ultracontractivity properties
of Feynman-Kac semigroups  for a class of L\'{e}vy processes", Annals of Probability, {\bf 43}, 1350, (2015).
\bibitem{lorinczi}  K. Kaleta and J. L\H{o}rinczi, "Transition in the decay rates  of stationary distributions of L\'{e}vy motion in an energy
landscape", Phys. Rev. E {\bf 93}, 022135, (2016).
\bibitem{simon} E. B. Davies and B. Simon, "Ultracontractivity and heat  kernels for Schr\"{o}dinger operators and Dirichlet
 Laplacians", J. Funct. Anal. {\bf 59},  335, (1984)
 \bibitem{acqua} Anna Dall'Acqua, "On the lifetime of a condtioned Brownian motion in the ball", J. Math. Anal. Appl. {\bf 335}, 389, (2007).
 \bibitem{bickel} T.  Bickel, "A note on the confined diffusion", Physica A {\bf 337}, 24, (2007)
 \bibitem{robinett} R. W. Robinett, "Visualizing the solutions of the circular infnite well in quantum and classical mechanics", Am. J. Phys. {\bf 64}, 440, (1996).
\bibitem{klauder} H. Ezawa, J. R. Klauder and L. A. Shepp, "A path space picture for Feynman-Kac averages", Ann. Phys. (NY), {\bf 88}, 588, (1974).
\bibitem{stef}  P. Garbaczewski and  V. Stephanovich, "L\'{e}vy flights and nonlocal quantum dynamics", J. Math. Phys. \textbf{54}, 072103, (2013).
\bibitem{stef0} P. Garbaczewski and V. Stephanovich, "L\'{e}vy flights in  confinig potentials", Phys. Rev. E  {\bf 80}, 031113, (2009).
\bibitem{stef1} P. Garbaczewski and  V. Stephanovich, "L\'{e}vy flights in inhomogeneous environments", Physica A {\bf 389}, 4419, (2010).
\bibitem{zaba} M. \.{Z}aba and P. Garbaczewski, "Solving fractional Schr\"{o}dinger-type spectral problems : Cauchy oscillator and Cauchy well",
J. Math. Phys. {\bf 55}, 092103, (2014).
\bibitem{zaba1} M. \.{Z}aba and P. Garbaczewski, "Nonlocally induced (fractional) bound states: Shape analysis
in the infnite Cauchy well", J. Math. Phys. {\bf 56}, 123502, (2015).
\bibitem{zaba2}  M. \.{Z}aba and P. Garbaczewski, "Ultrarelativistic bound states  in the spherical well",
 J. Math. Phys. {\bf 57}, 072302, (2016).
 \bibitem{zaba3} E. V.  Kirichenko., P. Garbaczewski, V. Stephanovich and M. \.{Z}aba, "L\'{e}vy flights in   an infnite well as a hypersingular Fredholm problem",
  Phys. Rev. E {\bf 93}, 052110, (2016).
\bibitem{lorinczi1} J. L\H{o}rinczi and J. Ma{\l}ecki, "Spectral properties of the massless relativistic harmonic oscillator", J. Diff. Equations,
{\bf 253},  2846, (2012)
\bibitem{lorinczi2} J. L\H{o}rinczi, K. Kaleta and S. O. Durugo, "Spectral and analytic properties of non-local Schr\'{o}dinger operators
and related jump processes", Comm. Appl. and Industrial Math. {\bf 6}(2), e-534, (2015);   DOI: 10.1685/journal.caim.534.
\bibitem{stos} T. Kulczycki, M. Kwa\'{s}nicki, J. Ma{\l}ecki, and A. St\'{o}s, "Spectral properties of the Cauchy process on half-line and interval",
Proc. London Math. Soc. {\bf 101}, 589–622 (2010).


\end{thebibliography}
\end{document}